\documentclass{aastex631}

\usepackage{graphicx}

\newcommand{\kms}{\ifmmode{\,\hbox{km\,s}^{-1}}\else
  {\rm\,km\,s$^{-1}$}\fi}
\newcommand{\masyr}{\rm\,mas\, yr$^{-1}~$}

\begin{document}


\title{The Multiple Extended Tidal Tails of NGC 288}

\author{Carl J. Grillmair}
\affil{IPAC, California Institute of Technology, Pasadena, CA 91125}
\email{carl@ipac.caltech.edu}

\begin{abstract}

  Using photometry and proper motions from Pan-STARRS, DECaLS, and
  Gaia DR3, we detect a $\sim 35\arcdeg$ to $70\arcdeg$-long trailing
  stellar debris stream associated with the globular cluster NGC
  288. The trajectory of the trailing tail is not well matched by a
  model stream evolved in a static Galactic potential, but is
  reasonably well-matched by a stream modeled in a potential that
  incorporates a massive, infalling Large Magellanic Cloud.  We also
  detect a broad, at least $\sim 40\arcdeg$-long leading tail that
  appears to be composed of at least two narrower, spatially offset,
  and kinematically distinct streams. Stream modeling predicts a
  similar broad composite of streams and suggests that these narrower
  components could each be made up of one or more generations of tidal
  tails, each formed during different orbits over the past few
  gigayears.On the other hand, NGC 288 is believed to have been
    brought into the Galactic halo during the Gaia-Enceladus-Sausage
    accretion event, and the tangential velocity dispersions of our
    stream candidates are indeed most consistent with having been
    stripped in a parent galaxy that had a large, cored dark matter
    halo. Tables of the most highly ranked stream star candidates are
  provided for ongoing and future spectroscopic surveys.

\end{abstract}



\section{Introduction}

Recent years have seen an enormous increase in the number of distinct
stellar debris streams detected in the Galactic halo
\citep{grillmair2016, shipp2018, malhan2018b, ibata2019, ibata2021,
  mateu2023}. A large number of these streams appear to be quite
narrow, with physical widths on the order of 100 pc, and are believed
to have been generated by globular clusters (GCs) formed {\it in situ}.  \citet{zhang2022} compiled a catalog of globular clusters
  with robust detections of tidal tails or extra-tidal
  structures. \citet{grondin2024} have recently computed positions and
  velocities of tidally stripped mock stars for all 159 globular
  clusters in the catalog of \citet{baumgardt2018}.
Thin, cold streams are particularly sensitive to interactions
with dark matter subhalos \citep{carlberg2009,yoon2011,erkal2016},
though episodic tidal stripping and associated variations in stream
density complicate the search for direct evidence of such interactions
\citep{kupper2012,thomas2016}.

On the other hand, by having well-characterized progenitors, tidal
tails associated with GCs will be useful for understanding the
detailed physics of tidal stripping \citep{balbinot2018}, the
accretion sequence of the halo, and the shape of the Galactic
potential \citep{bovy2016, price-whelan2014, malhan2019,
  garavito-camargo2021}.  Longer streams and streams on more eccentric
orbits are particularly sensitive to the shape of the halo potential
\citep{bonaca2018}.  \citet{carlberg2018} and \citet{malhan2021,
    malhan2022} showed that the spatial widths and velocity
  dispersions of GC tidal tails should be significantly larger for GCs
  that originated in and were initially stripped by infalling
  satellite galaxies with cored or cuspy cold dark matter halos. For
  all these reasons, it would be extremely useful to trace GC streams
  as deeply and as far around the Galaxy as the data will allow.

With accurate stellar proper motions over the entire sky, the third
data release (DR3) of the Gaia catalog \citep{gaiadr3} has made it
possible to detect stellar debris streams to far lower surface
densities than was possible with the purely photometric catalogs of
the pre-Gaia era \citep{malhan2018a, malhan2018b, grillmair2019,
  ibata2019, ibata2021, grillmair2022, yang2023}. In our
ongoing effort to extend individual globular cluster tidal streams as
far across the sky as possible, we focus here on the old, metal poor
globular cluster NGC 288.

The extended envelope and incipient tidal tails of NGC 288 were first
detected in the photographic work of \citet{grillmair1995} and were
among the strongest detections in their sample. \citet{leon2000} and
\citet{piatti2018} later found even more extensive extra-tidal
structures surrounding the cluster. \citet{shipp2018}, \citet{kaderali2019}, and
\citet{sollima2020} found a strong, wide plume of stars extending up
to $6\arcdeg$ to the south of NGC 288 and, most recently,
\citet{ibata2021} used STREAMFINDER \citep{malhan2018a} to detect a
narrow trailing tail extending some $8\arcdeg$ to the northwest of the
cluster.

In this work we combine Gaia DR3 with two ground-based photometric
catalogs to push the known limits of NGC 288's tidal tails even
further. Section \ref{analysis} describes our method, which largely
mirrors that used by \citet{grillmair2019} and \citet{grillmair2022}
to detect the extended tails of M5 and M2, respectively. We
  discuss the apparent trajectories , morphology, and dynamical widths
  of NGC 288's tails in Section \ref{discussion}. We also provide
  tables of our most highly ranked stream star candidates.  We make
  concluding remarks in Section \ref{conclusions}.

\section{Analysis} \label{analysis}

Our analysis generally follows that of \citet{grillmair2019}
(henceforth G19) and \citet{grillmair2022} (henceforth G22) as applied
to the globular clusters M5 and M2, combining color-magnitude and
proper motion filtering with orbit integrations and predictions based
on modeling the stripping of stars from NGC 288. We make use of the
photometry and proper motions contained in Gaia DR3 \citep{gaiadr3},
DECaLS \citep{dey2019}, and Pan-STARRS \citep{tonry2012, mast2022}.  We compute
the expected proper motions and trajectories on the sky using the
Galactic model of \citet{allen1991}, coded in the IDL language and
updated using the parameters of \citet{irrgang2013}. The model
includes a $6.6 \times 10^{10} M_\sun$ Miyamoto-Nagai disk
\citep{miyamoto1975}, a $9.5 \times 10^9 M_\sun$ spherical bulge, and
a $1.8 \times 10^{12} M_\sun$ spherical halo. Our experience using it
with other clusters (e.g. G19 and G22) suggests that it provides
reasonably good approximations for actual cluster orbits in the inner
halo.

It is now generally agreed \citep{kallivayalil2013, erkal2019, shipp2021}
that the Large Magellanic Cloud (LMC) is considerably more
massive than previously supposed and may be a significant perturber of
satellite orbits. Following G22, we augmented
our Galactic potential with a moving LMC, modeled as a point mass with
$M =1.88 \times 10^{11} M_\sun$ \citep{shipp2021} and  arriving at its
present location on a first pass. The LMC trajectory was modeled as a
fall from $\approx 700$ kpc, slowed by dynamical friction
\citep{chandrasekhar1943} according to the local density of stars and
dark matter along its orbit. While the stellar/dark matter wake of the
LMC is believed to be quite substantial \citep{conroy2021,
  garavito-camargo2021}, we have not attempted to model it as an
additional perturber on the orbit of NGC 288.

For modeling the orbit of NGC 288 and the motions of associated stream
stars, we used a cluster distance of $8.988 \pm 0.89$ kpc
\citep{baumgardt2021}, cluster proper motions of
$\mu_{\alpha} = 4.164 \pm 0.024$ \masyr and
$\mu_\delta = -5.705 \pm 0.024$ \masyr \citep{vasiliev2021}, and a
radial (line-of-sight) velocity of -44.83 km sec$^{-1}$
\citep{vasiliev2019}.  We computed the predicted trajectories of the
tidal tails by modeling the release of stars from the cluster over
time \citep{kupper2015, bowden2015, fardal2015, dai2018,
  grillmair2022}. Specifically, the model tails were generated by
releasing stars from the cluster at a rate proportional to $1/R^3$,
where $R$ is the Galactocentric radius of the cluster at any point in
time. Escaping stars were simply placed at the current $\approx 34$ pc
tidal radius of the cluster \citep{harris1996}, both at the L1 and L2
points, and then integrated independently along their individual
orbits. Orbit integrations were run for the equivalent of 5 Gyr to
provide tidal tails long enough to cover our field of interest. The
results of our modeling are shown in Figure 1.

 Also shown in Figure 1 is a simulation of stripped stars for NGC
  288 by \citet{grondin2024}.  While this simulation shows
  considerably more spread in all coordinates due to a longer
  integration time and to the inclusion of realistic three-body
  dynamics in the core of the cluster, there are identifiable tidal
  tails that match reasonably well with those in our models. There are
  small differences in the positional, velocity, and proper motion
  profiles at any given R.A., but this not unexpected given the
  different models used for the underlying Galactic potential. More
  importantly, the {\it shape} of the distance and proper motion
  profiles are very similar and, as we discuss below, we allow our
  distance and proper motion profiles to shift up and down to first
  discover and then to maximize the signal of stream stars.

\begin{figure}
\epsscale{1.0}
\plotone{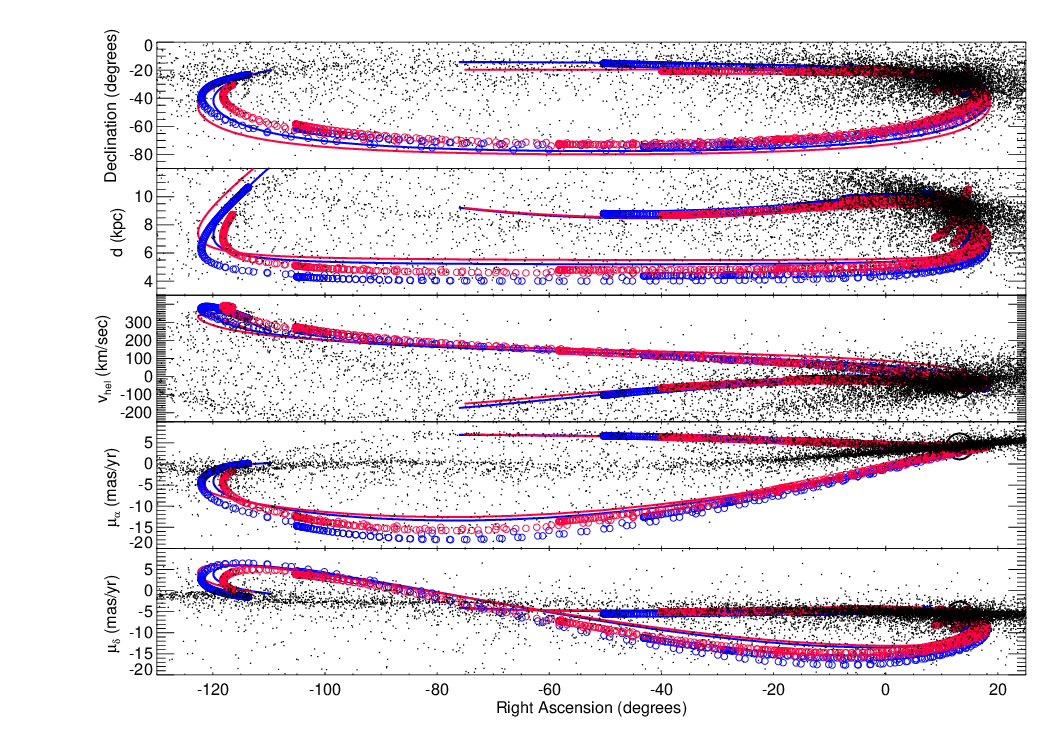}
\caption{The run of declination, distance, radial (line-of-sight) velocity, and proper
  motions with Right Ascension for the orbit of NGC 288 and its tidal
  tails, as predicted
  using the Galactic model of \citet{allen1991}, updated with model
  parameters from \citet{irrgang2013}, and with or without a
  perturbing LMC. The solid curves show orbit integrations of NGC
  288 itself, using just the \citet{allen1991} model (blue curve) and
  supplemented by a point-mass LMC fixed at its current position with
  $ M = 1.88 \times 10^{11} M_\sun $ (red curve). The small open circles
  show the results of sequentially releasing 2000 test particles from
  NGC 288's L1 and L2 locations and integrating their orbits separately. The blue circles
  are test particles integrated using just the Galactic model, while
  the red circles include the effect of a massive LMC falling from 700
  kpc to its current position on a first pass. The large black
  circles show the measured quantities for NGC 288 itself.  The black
  dots show the distributions of single stars in the tidal stripping
  simulation for NGC 288 by \citet{grondin2024}.}
\label{predictions}
\end{figure}

To reduce contamination by foreground stars, we limited our analysis
to stars with $15 < G < 20.0$. For stars fainter than $G = 20$, the
Gaia photometry and proper motion uncertainties are large enough
that their inclusion results in noticeably noisier results.

We used Gaia, Pan-STARRS, and DECaLS photometry to determine
photometric membership probabilities for individual stars. We used a
modified form of the matched filter described by \citet{rockosi2002}
and \citet{grillmair2009}. Specifically, we weighted stars both by
their position in the extinction-corrected $G_0, (G_{BP} - G_{RP})_0$,
$G_0, (g_{PS1} - r_{PS1})_0$, $G_0, (g_{PS1} - i_{PS1})_0$,
$G_0, (g_{Decam}-r_{Decam})_0$, and $G_0, (g_{Decam}-i_{Decam})_0$
color-magnitude diagrams of NGC 288 using an error-weighted
average. The $G$ magnitudes were adjusted faintward or brightward by
sky location to account for the predicted variation in the
heliocentric distances of stars along the leading and trailing
arms. Photometry was corrected for extinction using the reddening maps
of \citet{schlegel1998}, themselves corrected using the prescription
of \citet{schlafly2011}, and using the Gaia DR2 coefficients derived
by \citet{gaia2018}, the Pan-STARRS coefficients provided by
\citet{tonry2012}, and the DECaLS coefficients given at
\url{https://www.legacysurvey.org/dr5/description/}.

We also weighted stars by their departures from the predicted proper
motion profiles shown in Figure 1.  Since the proper motions and
heliocentric distances at any given R.A. are not unique, we
imposed the predicted trajectories of the leading and
trailing arms on the sky so that the photometric and proper motion filtering applied
to any particular star would depend on which point along which arm it is closest to.

While Gaia DR3 parallax measurements are generally not very accurate
at the distance of NGC 288, we carried out experiments using parallax
as an additional constraint to weed out both foreground dwarfs and very
distant stars or galaxies. These experiments showed a small but
noticeable improvement in signal-to-noise ratio. All results shown here
consequently employed an additional weighting factor based on the
expected parallax at each point along each arm.

Stream membership probabilities were computed as a product of the CMD,
$\mu_\alpha$, $\mu_\delta$, and parallax membership probabilities. We
then summed the resulting filter signals by sky position to produce
probability surface density maps we could examine for evidence of
tidal streams. Several thousand orbit models were computed over a
three-dimensional grid of distance, $\mu_{\alpha}$, and
$\mu_{\delta}$. While the proper motions at every point along the
tidal tails predicted by our relatively simple model of the Galactic
potential are not terribly accurate, the overall {\it shape} of the
proper motion profiles in Figure 1 are reasonably invariant over the
range of uncertainties in NGC 288's distance, proper motion, and
radial velocity. We consequently shifted the proper motion profiles up
and down by up to 5 \masyr in increments of 0.1 \masyr.  Similarly,
the distance profile in Figure 1 was shifted up and down by up to 1 kpc
in increments of 200 pc. The resulting surface density maps were then
examined by eye to look for obvious enhancements along the expected
positions of the tidal tails.

We restricted our search to the Southern Galactic cap. Our models
predict that NGC 288's tidal arms do not extend much beyond
$10\arcdeg$ north of the Galactic plane. Moreover, the proper motions
of tail stars north of the Galactic plane are predicted to be very
close to zero and virtually indistinguishable from those of the field
population.

\section{Discussion} \label{discussion}

Figure 2 shows clear plumes of stars emmanating from NGC 288, both
along the $l = 310\arcdeg$ meridian, and in the general direction of
the $l = 130\arcdeg$ meridian. The northern portion of the very
strong, $10\arcdeg$-long leading plume extending along the
$l = 310\arcdeg$ meridian was detected by \citet{shipp2018} in Dark
Energy Survey data and by \citet{kaderali2019} and \citet{sollima2020}
in Gaia DR2. We refer to this feature as a plume (as distinct from a
tidal tail) owing to its relative compactness, its $\sim 40\arcdeg$
misalignment with the expected orbit of the cluster and the track of
the stream, and its similarity to relatively recent ejections of stars
apparent in Figure 1. The stars in such plumes eventually spread out
along their own orbits, creating new tidal tails that may or may not
be easily distinguishable from previous generations
\citep{hozumi2015}.

\begin{figure}
\epsscale{1.0}
\includegraphics[width=7in]{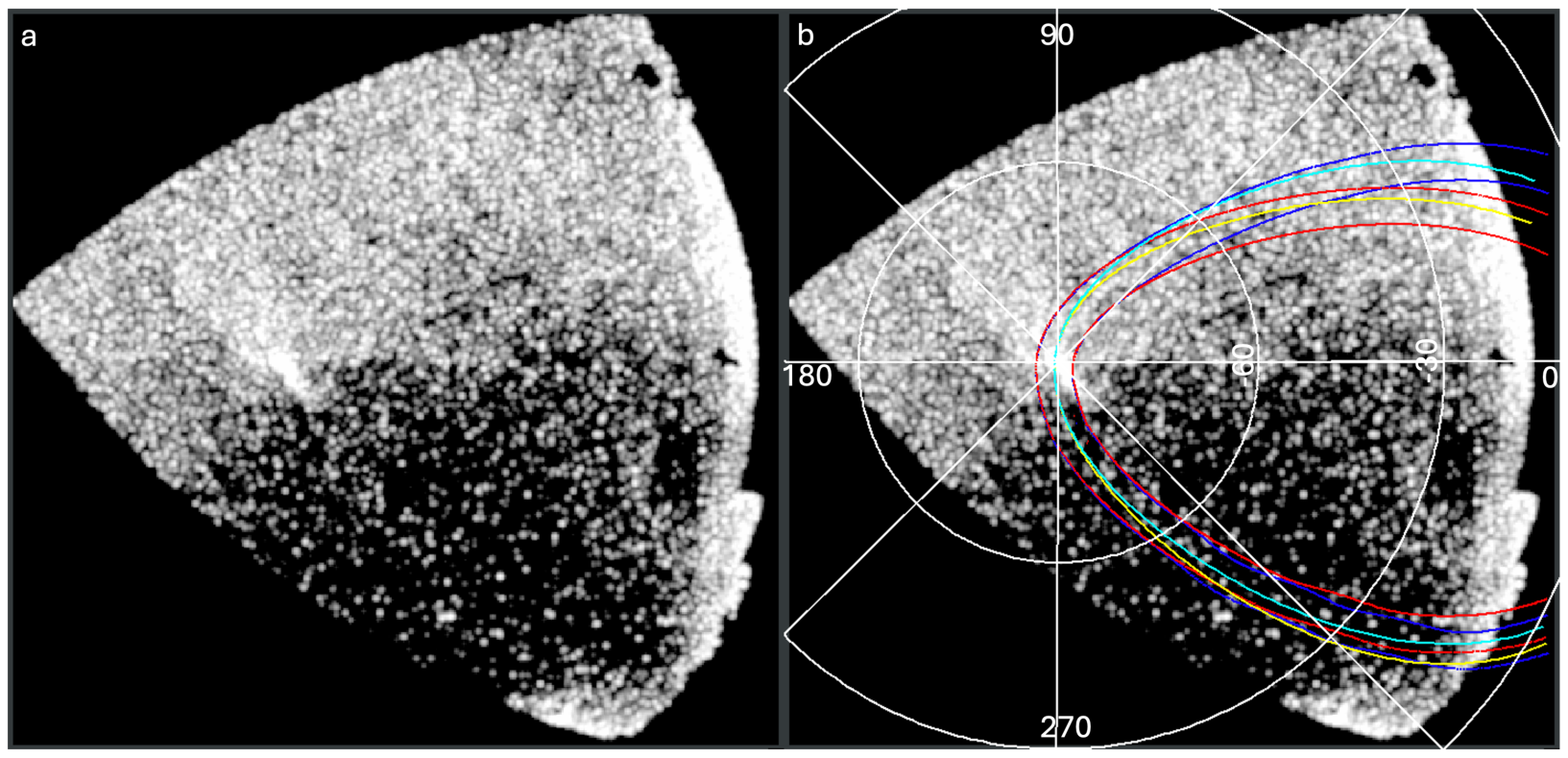}
\caption{Probability surface density maps of NGC 288 and its tidal
  arms. This polar projection, centered on the south Galactic pole,
  encompasses $\approx 10,000$ square degrees, or roughly one quarter
  of the sky. The stretch is logarithmic, the scale is $0.3\arcdeg$
  per pixel, and the maps have been smoothed with a Gaussian kernel of
  $0.3\arcdeg$. The background discontinuities across the middle of
  the images are a consequence of applying different proper motion
  filters to areas near the leading and trailing tidal arms,
  respectively. Panel (b) is identical in scale and stretch to panel
  (a) but provides a Galactic coordinate grid along with an overlay of
  the cluster orbits and the model streams shown in Figure 1. The
  yellow and cyan curves show the orbit of NGC 288 itself, with and
  without a massive LMC, respectively. The red and blue curves are
  offset $3\arcdeg$ from the centerlines of tidal tail models with and
  without a massive infalling LMC, respectively.}
\label{map}
  \end{figure}

  To the upper left in Figure 2, the trailing plume is much broader,
  extending at least $20\arcdeg$ from the cluster in a fan-shaped
  distribution. These stars are only slightly separated from one
  another in proper motion, and a sweep through proper motion space
  highlights different portions of the plume monotonically from one
  side to the other.

\subsection{Leading Tail}

There is a clear surplus of stars (detected at the $\sim 4\sigma$ level
with respect to field stars using the T-statistic of \citet{grillmair2009})
extending at least $40\arcdeg$ north of the cluster along the
$l \sim 300\arcdeg$ meridian, and possibly up to $\approx 80\arcdeg$
from the cluster, where it becomes overwhelmed by the Galactic
disk. This feature, which we will refer to as NGC 288's leading tail,
is in good accord with the trajectories of the leading tails predicted by
models both with and without a massive, infalling LMC.

At roughly $5\arcdeg$ across, the leading tail is quite broad. From
Figure 1 we see that the leading tail passes within 6 kpc of the Sun
on its way to perigalacticon. At this distance an apparent width of
$5\arcdeg$ would correspond to about 500 pc. This would be unusually
broad compared with other presumed globular cluster streams ({\it
  e.g.} GD-1 and Pal 5). However, careful examination of the surface
density maps in the 3-dimensional grid of $\mu_{\alpha}$,
$\mu_{\delta}$, and distance shows that this leading tail is made up of 
at least two components, separated in proper motion by $\sim 1$ mas
yr${^{-1}}$ in $\mu_{\alpha}$ and $\sim 0.3$ mas yr$^{-1}$ in
$\mu_{\delta}$. Figure 3 shows surface density maps generated after
appropriately offsetting the proper motions from the nominal values used for Figure
2. Each panel shows a $\approx 5\sigma$ excess of stars over the
background, and the visible streams are laterally offset from one another by
about $4\arcdeg$. The apparent widths of these sub-streams are each less than
$2\arcdeg$, or about 200 pc.

\begin{figure}
\epsscale{1.0}
\includegraphics[angle=270,width=7in]{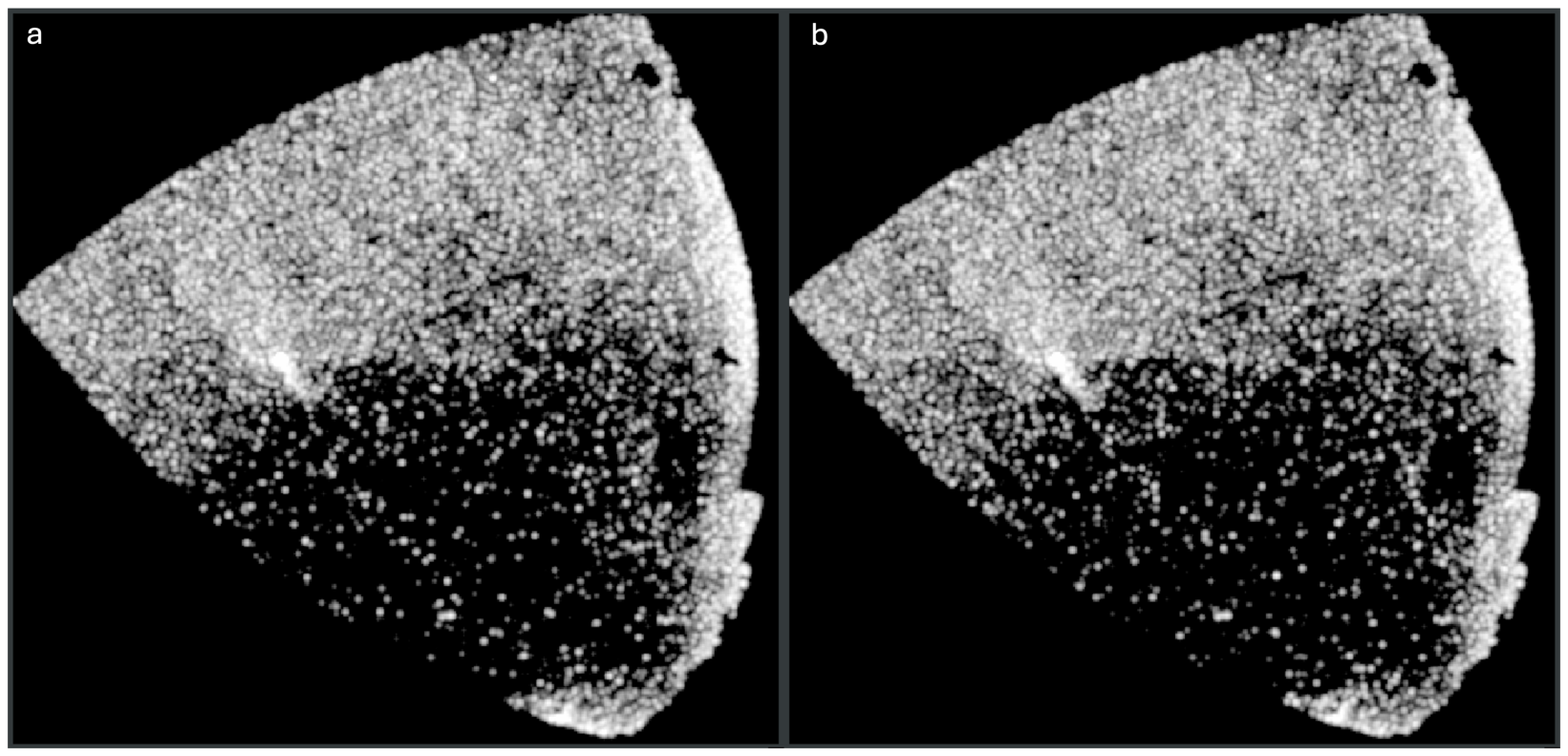}
\caption{Same as Figure 2, but using proper motion profiles offset
  from one another by 1 \masyr in $\mu_{\alpha}$ and 0.3 \masyr in
  $\mu_{\delta}$. Panel (a) shows a probability map computed for the
  ``outer'' leading stream, while panel (b) shows a map computed for
  the ``inner'' leading stream.}
\label{leading_tail_maps}
  \end{figure}

\begin{figure}
\epsscale{1.0}
\plotone{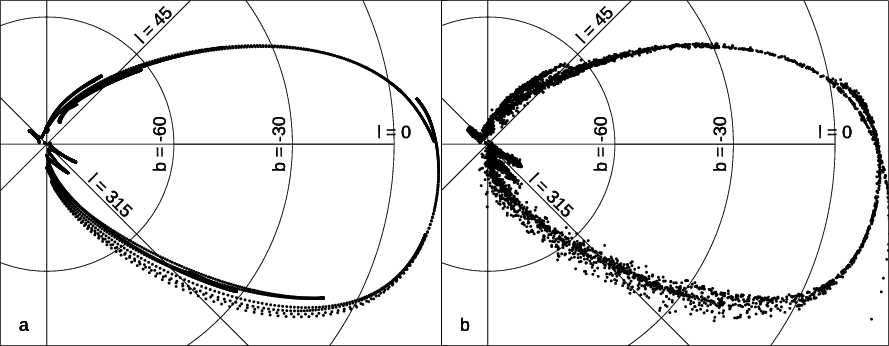}
\caption{A model of NGC 288's tidal tails after running for the
  equivalent of 3 Gyr in a Galactic potential that incorporates a massive infalling
LMC. Panel (a) shows a model in which stripped stars have zero
velocity dispersion at the time of release. Panel (b) shows results
for the same model, but with stars that are given a velocity dispersion of 1
km s$^{-1}$ at the time of release. The projection and
orientation are identical to those used in Figures \ref{map} and \ref{leading_tail_maps}.}
\label{models}
  \end{figure}

  For comparison, we show a 5000-particle model, run forward from a
  point 3 Gyr in the past, and including a massive infalling LMC in
  Figure \ref{models}. At $b = -60\arcdeg$, the different generations
  of leading tails produce a swath more than $5\arcdeg$ across, in
  good agreement with the observed tail in Figure 2. The individual
  tails do not arrange themselves in any particular order; there are
  both older and younger generations of tails populating both the
  inner and outer regions of the swath. The formation of distinct
  tidal tails during each orbit has been examined in greater detail by
  \citet{montuori2007} and \citet{hozumi2015}, among others. 
    Though the physical widths of tidal tails are known to depend on
    the size and concentration of the galaxies in which a cluster
    first formed \citep{carlberg2018,malhan2021}, our modeling
    suggests that the overall width of NGC 288's leading tail could 
    be a natural consequence of the precession of the cluster's
    orbit in the apsherical potential of the Galaxy.

  Averaging over the inner (western) and outer (eastern) streams in
  the model, we find that the proper motion offsets between the
  substreams in $\mu_{\alpha}$ and $\mu_{\delta}$ are about 0.2 mas
  yr$^{-1}$ and 1.1 mas yr$^{-1}$, respectively. While similar in
  total magnitude to what we find observationally, these offsets are
  reversed in the sense that $\mu_{\alpha}$ is the larger component of
  the offset. We attribute this disagreement to the limited fidelity
  of our model potential which, among many other things, does not
  include a bar or a Galactic wake due to perturbations by the
  LMC.  Comparing photometric distances for candidate stars in the
    southern $40\arcdeg$ of each stream (see below), we find a mean
    distance offset of $93 \pm 110$ pc, in the sense that the outer
    (western) stream may be slightly closer to us than the inner
    (eastern) stream. Though consistent with zero, this offset agrees
    with with a $\approx 75$ pc offset, in the same sense, between the
    western and eastern halves of the model stream in Figure
    \ref{models}.  Combined, these offsets support the notion that we
  are seeing multiple generations of tidal tails. While each of the
  streams visible in Figure 3 are likely made up of stars stripped
  during multiple orbits, for simplicity we henceforth refer to the
  stream in panel (a) of Figure 3 as the outer leading stream, and the
  stream in panel (b) as the inner leading stream.

  \subsection{Trailing Tail}

  NGC 288's trailing tail extends to the upper right of the cluster in
  Figure 2 roughly along $l = 55\arcdeg$. The first $8\arcdeg$ of this
  tail was detected by \citet{ibata2021}. The extended trailing tail in Figure
  2, while more contaminated by foreground stars
  with similar proper motions than the leading tail,  is detectable to
  $b \sim -55\arcdeg$. Beyond that the signal becomes more tenuous,
  essentially disappearing in the region
  $-58\arcdeg < b < -48\arcdeg$, and apparently resurging in the
  region $-40\arcdeg < b < -25\arcdeg$. Over the interval
  $-90\arcdeg < b < -25\arcdeg$ the stream is detected at the
  $4.7\sigma$ level. The trailing tail appears somewhat narrower on the
  sky than the leading tail, with a full-width-at-half-maximum of
  about $1.5\arcdeg$. At a distance of between 8.6 and 9.4 kpc, this
  corresponds to a physical width of between 200 and 250
  pc. Interestingly, this narrower profile is also reflected in the
  model tails in Figure \ref{models}, with different generations of
  trailing tails nearly overlapping from our Galactic vantage point.

The trailing tail is reasonably well matched in position by a stream
model incorporating a massive infalling LMC (Figure 2). On the other hand, the
trajectory predicted by a model without an infalling LMC departs
fairly quickly from that of the observed stream. This is qualitatively
similar to the results of G22 for NGC 7089, who found that both the
trajectory and the proper motions of NGC 7089's observed trailing tail
were better matched by a model incorporating a massive infalling LMC.

\subsection{Candidate Tail Stars}

\begin{figure}
\epsscale{0.5}
\includegraphics[angle=270,width=7in]{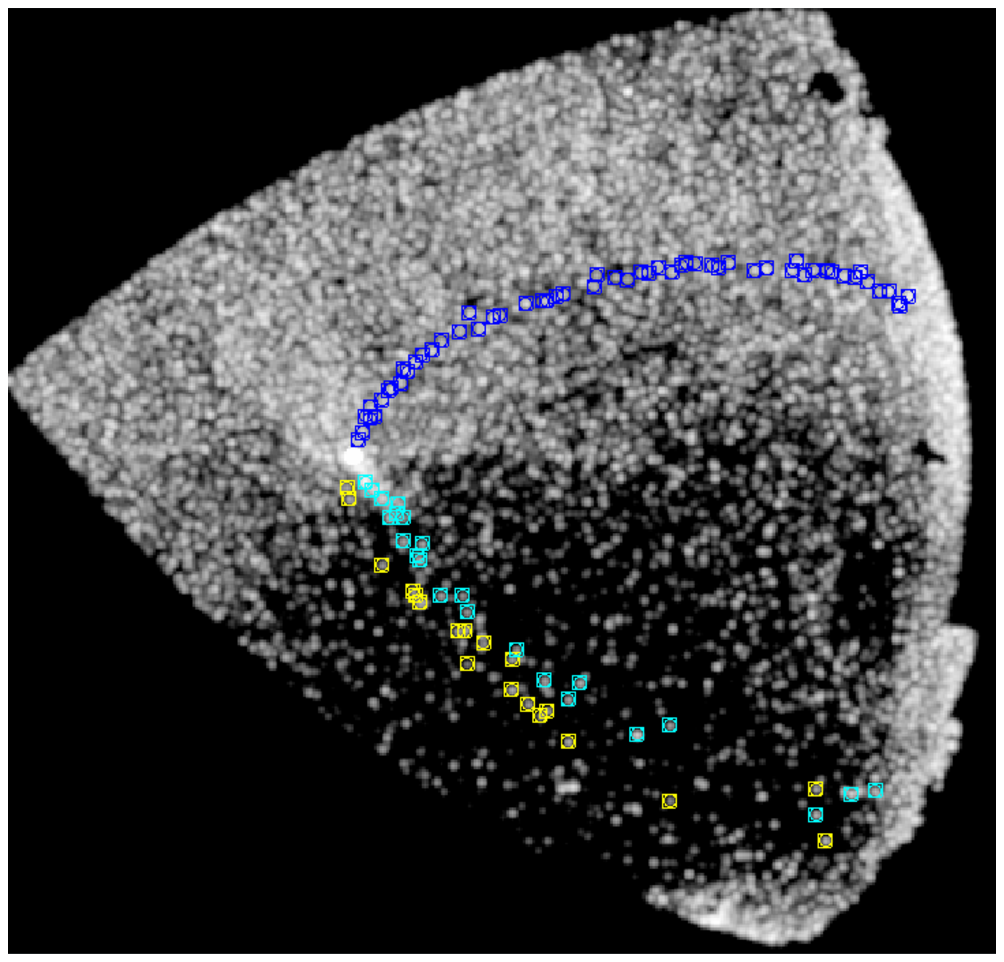}
\caption{High-probability members of NGC 288's tidal tails. The blue
  points show stars lying along the trailing tail, the cyan points
  designate candidate stars in the inner leading tail, and yellow
  points show stars we assign to the outer leading tail.}
\label{map_of_selected}
\end{figure}

\begin{figure}
\epsscale{1.0}
\includegraphics{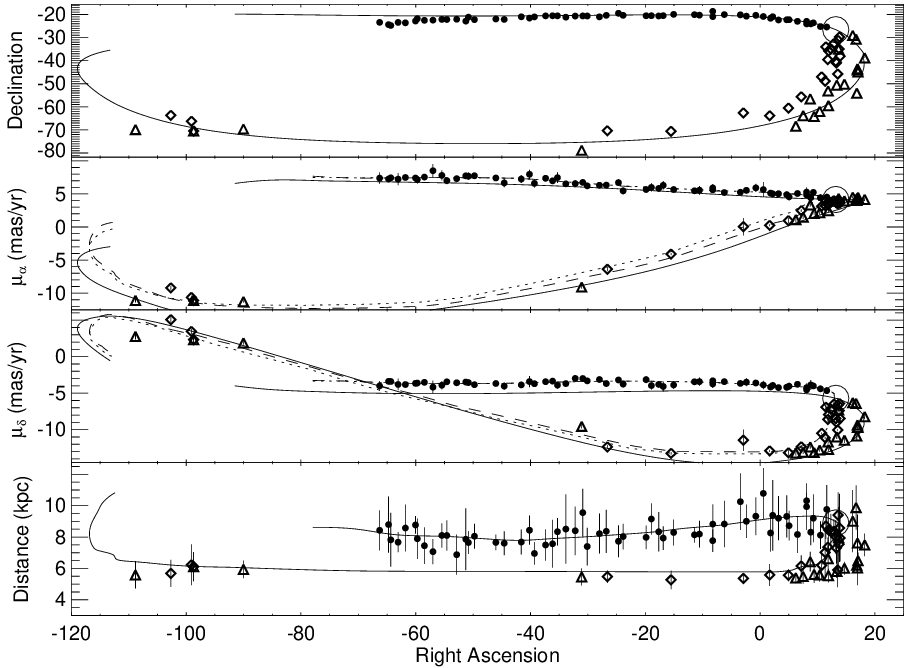}
\caption{ Positions, proper motions, and photometrically-estimated
  distances of our highest-ranked candidate tail stars, as listed in
  Tables 1, 2, and 3. The open circles denote NGC 288 itself. The
  filled circles show candidates in our trailing tail, while open
  triangles and open diamonds show candidates in our inner and outer
  leading tails, respectively.  The solid curve is our nominal model of
  the positions, proper motions, and distances of stars in the tails based on our
  adopted Galactic model and a massive infalling LMC. The dashed and
  dotted curves are the offset proper motion profiles used to maximize
  the signal for the inner and outer streams, respectively.}
\label{positions}
  \end{figure}

Gaia DR3 radial velocities extend to $G_{RVS} \approx 14$
\citep{katz2023}, which unfortunately is considerably brighter than
our brightest high-probability candidate ($G = 17.8$).  A positional
matching of our highest-weighted stars with the Sloan Digital Sky
Survey Data Release 15 \citep{aguado2018} as well as the LAMOST Data
Release 6 \citep{zhao2012} did not yield any radial velocity
measurements. To enable radial velocity follow-up measurements by the
community, Tables 1, 2, and 3 list stars with the highest filter
signals and lying along the trailing, outer leading, and inner leading
tails, respectively. Each table is sorted by increasing Galactic
latitude. Only stars with membership probabilities exceeding 0.22
  (which would correpond to a star $1\sigma$ from the expected
  color-magnitude locus and $1\sigma$ from each of the expected
  $\mu_\alpha$ and $\mu_\delta$ profiles at any position along the
  stream) are listed in the tables.  Distances are estimated by fixing
  the proper motion profiles at their ``best'' values, running the
  analysis over a range of assumed distances and determining the
  distance at which agreement between a star's color and the expected
  color-magnitude locus at that sky position is maximized.  Distance
  uncertainties are estimated by noting the distances at which the
  maximum CMD membership probability drops by a factor of 0.607 and
  thus correspond roughly to $1\sigma$ errors. In instances
  where stars could be either main sequence or subgiant stars, we
  choose the distance that best matches expectations for stream stars
  at that sky location, though the uncertainty is increased 
  to reflect this degeneracy. Both leading and trailing tails become
almost indistinguishable from the background near the Galactic plane.
The positions, proper motions, and distances of all tabulated stars
are plotted in Figures \ref{map_of_selected} and
\ref{positions}.

\startlongtable
\begin{deluxetable*}{rcccccccc}
\tabletypesize{\small}
\tablecaption{Candidate Stream Stars: Trailing Tail}
\tablecolumns{9}

\tablehead{ 
\colhead{No.} & \colhead{R.A. (J2016)} & \colhead{dec (J2016)} & \colhead{$G$} & \colhead{$G_{BP} - G_{RP}$} & \colhead{$\mu_\alpha \cos{\delta}$ (mas yr$^{-1}$)}& \colhead{$\mu_\delta$ (mas yr$^{-1}$)} & \colhead{Distance (kpc)} & \colhead{Probability}
}

\startdata
  1 &     11.6178 &    -25.4231 &  19.846 &  0.849 &   4.519 $\pm$   0.500 &  -4.652 $\pm$   0.444 &   9.76 $\pm$   1.56 &   0.78 \\ 
  2 &     10.5210 &    -25.2142 &  19.778 &  0.846 &   4.433 $\pm$   0.481 &  -4.402 $\pm$   0.405 &   8.12 $\pm$   0.88 &   0.40 \\ 
  3 &      9.2878 &    -23.6151 &  19.271 &  0.686 &   5.240 $\pm$   0.369 &  -4.006 $\pm$   0.248 &   9.20 $\pm$   1.52 &   0.34 \\ 
  4 &      8.8291 &    -24.0366 &  20.015 &  0.619 &   4.435 $\pm$   0.582 &  -3.868 $\pm$   0.659 &   8.31 $\pm$   0.83 &   0.36 \\ 
  5 &      8.1098 &    -24.0821 &  18.781 &  0.693 &   4.847 $\pm$   0.205 &  -4.534 $\pm$   0.261 &  10.31 $\pm$   0.50 &   0.24 \\ 
  6 &      8.1033 &    -22.9501 &  19.944 &  0.773 &   5.287 $\pm$   0.698 &  -4.241 $\pm$   0.598 &   9.94 $\pm$   1.48 &   0.76 \\ 
  7 &      6.5142 &    -22.9660 &  19.231 &  0.668 &   5.068 $\pm$   0.336 &  -3.781 $\pm$   0.254 &   8.16 $\pm$   1.18 &   0.35 \\ 
  8 &      5.0590 &    -22.2461 &  19.979 &  0.735 &   4.564 $\pm$   0.543 &  -4.037 $\pm$   0.461 &   8.72 $\pm$   0.94 &   0.51 \\ 
  9 &      4.6430 &    -22.2799 &  19.821 &  0.712 &   4.894 $\pm$   0.451 &  -4.184 $\pm$   0.440 &   9.32 $\pm$   1.34 &   0.82 \\ 
 10 &      3.2104 &    -22.3650 &  19.018 &  0.666 &   5.036 $\pm$   0.332 &  -4.233 $\pm$   0.204 &   9.20 $\pm$   1.41 &   0.42 \\ 
 11 &      1.8234 &    -21.6370 &  19.543 &  0.896 &   5.129 $\pm$   0.389 &  -4.156 $\pm$   0.297 &   8.26 $\pm$   1.19 &   0.59 \\ 
 12 &      2.2048 &    -20.9640 &  19.054 &  0.731 &   5.021 $\pm$   0.294 &  -3.937 $\pm$   0.245 &   9.38 $\pm$   2.25 &   0.82 \\ 
 13 &      0.5652 &    -20.8605 &  19.811 &  0.668 &   5.664 $\pm$   1.106 &  -3.868 $\pm$   0.565 &  10.78 $\pm$   1.62 &   0.55 \\ 
 14 &    359.2597 &    -20.3508 &  19.864 &  0.775 &   5.923 $\pm$   0.602 &  -3.600 $\pm$   0.378 &   9.33 $\pm$   1.20 &   0.29 \\ 
 15 &    357.6157 &    -20.7251 &  19.921 &  0.882 &   5.545 $\pm$   0.468 &  -3.488 $\pm$   0.544 &   9.00 $\pm$   1.06 &   0.54 \\ 
 16 &    356.5639 &    -20.1499 &  19.467 &  0.880 &   5.259 $\pm$   0.327 &  -3.663 $\pm$   0.344 &  10.25 $\pm$   1.80 &   0.75 \\ 
 17 &    353.7707 &    -20.0313 &  19.243 &  0.679 &   5.212 $\pm$   0.304 &  -3.416 $\pm$   0.334 &   8.83 $\pm$   1.47 &   0.68 \\ 
 18 &    351.7619 &    -20.9221 &  18.938 &  0.639 &   5.619 $\pm$   0.360 &  -3.305 $\pm$   0.278 &   8.83 $\pm$   1.26 &   0.41 \\ 
 19 &    351.7529 &    -18.6240 &  19.999 &  0.632 &   5.910 $\pm$   0.573 &  -3.725 $\pm$   0.552 &   7.77 $\pm$   0.72 &   0.36 \\ 
 20 &    349.4342 &    -20.2762 &  19.877 &  0.857 &   5.560 $\pm$   0.644 &  -3.417 $\pm$   0.565 &   8.19 $\pm$   0.85 &   0.79 \\ 
 21 &    348.5678 &    -20.0669 &  17.930 &  0.902 &   5.478 $\pm$   0.129 &  -3.438 $\pm$   0.116 &   8.14 $\pm$   0.49 &   0.35 \\ 
 22 &    344.9409 &    -19.9379 &  19.995 &  0.819 &   5.670 $\pm$   0.503 &  -3.871 $\pm$   0.520 &   8.29 $\pm$   0.83 &   0.60 \\ 
 23 &    343.1269 &    -20.4930 &  20.020 &  0.981 &   6.307 $\pm$   0.622 &  -4.078 $\pm$   0.664 &   7.93 $\pm$   0.75 &   0.34 \\ 
 24 &    342.3369 &    -20.6402 &  19.578 &  0.755 &   5.842 $\pm$   0.483 &  -3.792 $\pm$   0.399 &   8.33 $\pm$   1.25 &   0.55 \\ 
 25 &    341.0597 &    -20.5189 &  20.060 &  0.704 &   5.989 $\pm$   0.637 &  -3.954 $\pm$   0.554 &   9.15 $\pm$   1.00 &   0.38 \\ 
 26 &    340.1381 &    -20.5307 &  20.056 &  0.870 &   5.674 $\pm$   0.495 &  -3.143 $\pm$   0.490 &   7.98 $\pm$   0.75 &   0.49 \\ 
 27 &    336.1360 &    -20.3584 &  19.987 &  0.830 &   5.469 $\pm$   0.524 &  -3.748 $\pm$   0.468 &   8.03 $\pm$   0.81 &   0.24 \\ 
 28 &    335.3248 &    -19.4792 &  19.796 &  0.871 &   6.716 $\pm$   0.475 &  -3.188 $\pm$   0.400 &   7.73 $\pm$   1.01 &   0.37 \\ 
 29 &    333.2042 &    -20.3747 &  19.604 &  0.846 &   6.342 $\pm$   0.336 &  -3.682 $\pm$   0.318 &   8.36 $\pm$   1.37 &   0.52 \\ 
 30 &    332.0131 &    -21.0193 &  18.772 &  0.662 &   6.316 $\pm$   0.292 &  -3.124 $\pm$   0.209 &   8.22 $\pm$   1.27 &   0.39 \\ 
 31 &    329.8831 &    -20.6390 &  19.064 &  0.701 &   6.226 $\pm$   0.291 &  -3.307 $\pm$   0.287 &   7.40 $\pm$   1.19 &   0.41 \\ 
 32 &    329.1082 &    -20.6083 &  19.098 &  0.603 &   6.747 $\pm$   0.299 &  -2.988 $\pm$   0.339 &   9.55 $\pm$   1.51 &   0.24 \\ 
 33 &    327.7895 &    -20.6783 &  19.401 &  0.745 &   6.568 $\pm$   0.517 &  -2.969 $\pm$   0.389 &   8.40 $\pm$   1.55 &   0.44 \\ 
 34 &    326.1528 &    -21.1658 &  18.878 &  0.700 &   6.532 $\pm$   0.300 &  -3.684 $\pm$   0.253 &   8.51 $\pm$   2.21 &   0.39 \\ 
 35 &    324.6987 &    -20.5853 &  19.966 &  0.624 &   7.471 $\pm$   0.791 &  -3.897 $\pm$   0.462 &   8.34 $\pm$   0.87 &   0.48 \\ 
 36 &    323.8435 &    -20.5527 &  18.646 &  0.717 &   6.983 $\pm$   0.224 &  -3.342 $\pm$   0.164 &   7.57 $\pm$   1.52 &   0.74 \\ 
 37 &    322.6260 &    -20.5994 &  19.182 &  0.772 &   7.369 $\pm$   0.376 &  -3.424 $\pm$   0.280 &   7.49 $\pm$   0.99 &   0.32 \\ 
 38 &    320.6786 &    -21.0356 &  19.617 &  0.936 &   6.599 $\pm$   0.625 &  -3.702 $\pm$   0.360 &   6.95 $\pm$   0.72 &   0.29 \\ 
 39 &    319.8177 &    -21.5507 &  19.914 &  0.874 &   7.968 $\pm$   0.672 &  -3.411 $\pm$   0.747 &   8.43 $\pm$   0.93 &   0.34 \\ 
 40 &    318.3910 &    -20.9930 &  19.765 &  0.765 &   7.257 $\pm$   0.582 &  -3.867 $\pm$   0.411 &   7.68 $\pm$   0.79 &   0.75 \\ 
 41 &    315.4156 &    -22.1397 &  19.953 &  0.977 &   6.672 $\pm$   0.625 &  -4.029 $\pm$   0.471 &   7.61 $\pm$   0.75 &   0.34 \\ 
 42 &    313.9588 &    -21.9383 &  18.394 &  0.806 &   7.417 $\pm$   0.190 &  -3.670 $\pm$   0.160 &   7.66 $\pm$   0.50 &   0.82 \\ 
 43 &    310.2284 &    -22.2375 &  19.955 &  0.706 &   7.753 $\pm$   0.436 &  -3.834 $\pm$   0.356 &   8.05 $\pm$   0.79 &   0.73 \\ 
 44 &    308.7216 &    -22.8732 &  18.726 &  0.714 &   7.743 $\pm$   0.251 &  -3.472 $\pm$   0.185 &   7.86 $\pm$   1.97 &   0.35 \\ 
 45 &    309.1761 &    -21.1555 &  20.015 &  0.903 &   7.688 $\pm$   0.564 &  -3.547 $\pm$   0.387 &   7.65 $\pm$   0.72 &   0.77 \\ 
 46 &    307.1371 &    -22.0531 &  18.518 &  0.760 &   7.315 $\pm$   0.206 &  -3.571 $\pm$   0.145 &   6.88 $\pm$   1.27 &   0.47 \\ 
 47 &    305.4191 &    -22.1641 &  19.826 &  0.801 &   7.012 $\pm$   0.515 &  -3.437 $\pm$   0.359 &   8.10 $\pm$   0.97 &   0.52 \\ 
 48 &    304.5562 &    -22.1268 &  20.201 &  1.009 &   7.805 $\pm$   0.766 &  -3.909 $\pm$   0.578 &   8.10 $\pm$   0.78 &   0.84 \\ 
 49 &    302.9991 &    -22.4355 &  20.342 &  1.101 &   8.514 $\pm$   1.003 &  -4.168 $\pm$   0.875 &   7.07 $\pm$   0.80 &   0.27 \\ 
 50 &    301.5133 &    -22.5263 &  19.752 &  1.095 &   7.557 $\pm$   0.440 &  -3.514 $\pm$   0.305 &   7.45 $\pm$   1.17 &   0.75 \\ 
 51 &    299.9524 &    -22.8098 &  18.765 &  0.767 &   7.415 $\pm$   0.227 &  -3.652 $\pm$   0.153 &   8.77 $\pm$   0.56 &   0.78 \\ 
 52 &    300.2743 &    -21.8105 &  19.920 &  0.842 &   7.161 $\pm$   0.589 &  -3.612 $\pm$   0.327 &   7.89 $\pm$   1.28 &   0.83 \\ 
 53 &    298.1937 &    -23.6058 &  19.271 &  0.854 &   7.490 $\pm$   0.285 &  -3.815 $\pm$   0.177 &   8.58 $\pm$   1.50 &   0.44 \\ 
 54 &    296.9223 &    -23.4210 &  20.053 &  0.715 &   7.239 $\pm$   0.917 &  -3.806 $\pm$   0.791 &   7.67 $\pm$   0.64 &   0.70 \\ 
 55 &    295.6508 &    -24.7409 &  18.670 &  0.795 &   7.444 $\pm$   0.213 &  -3.393 $\pm$   0.205 &   7.82 $\pm$   1.71 &   0.78 \\ 
 56 &    295.2555 &    -24.3674 &  18.872 &  0.755 &   7.262 $\pm$   0.214 &  -3.381 $\pm$   0.197 &   8.79 $\pm$   1.78 &   0.38 \\ 
 57 &    293.7054 &    -23.4103 &  20.142 &  0.882 &   7.395 $\pm$   0.885 &  -4.003 $\pm$   0.663 &   8.43 $\pm$   1.28 &   0.71 \\ 
\enddata
\end{deluxetable*}

\startlongtable
\begin{deluxetable}{rcccccccc}
\tabletypesize{\small}
\tablecaption{Candidate Stream Stars: Outer Leading Tail}
\tablecolumns{9}

\tablehead{ 
\colhead{No.} & \colhead{R.A. (J2016)} & \colhead{dec (J2016)} & \colhead{$G$} & \colhead{$G_{BP} - G_{RP}$} & \colhead{$\mu_\alpha \cos{\delta}$ (mas yr$^{-1}$)} & \colhead{$\mu_\delta$ (mas yr$^{-1}$)} & \colhead{Distance (kpc)} & \colhead{Probability}
}
\startdata
  1 &     16.1455 &    -29.0652 &  18.971 &  0.719 &   4.515 $\pm$   0.192 &  -6.301 $\pm$   0.318 &   9.01 $\pm$   1.97 &   0.29 \\ 
  2 &     16.7313 &    -30.6763 &  19.696 &  0.758 &   4.180 $\pm$   0.340 &  -6.386 $\pm$   0.233 &   9.88 $\pm$   1.41 &   0.43 \\ 
  3 &     13.6871 &    -35.0034 &  18.676 &  0.689 &   4.318 $\pm$   0.120 &  -7.297 $\pm$   0.127 &   7.96 $\pm$   1.97 &   0.58 \\ 
  4 &     18.2333 &    -38.7918 &  19.679 &  0.775 &   4.176 $\pm$   0.213 &  -8.204 $\pm$   0.301 &   7.50 $\pm$   0.71 &   0.38 \\ 
  5 &     16.9770 &    -43.4823 &  18.527 &  0.715 &   4.350 $\pm$   0.097 &  -9.337 $\pm$   0.106 &   7.62 $\pm$   1.81 &   0.47 \\ 
  6 &     16.9643 &    -43.9691 &  18.113 &  0.669 &   4.315 $\pm$   0.078 &  -9.680 $\pm$   0.098 &   6.04 $\pm$   1.09 &   0.52 \\ 
  7 &     17.0685 &    -44.8767 &  19.737 &  0.824 &   4.495 $\pm$   0.220 &  -9.655 $\pm$   0.297 &   6.49 $\pm$   0.51 &   0.61 \\ 
  8 &     14.6984 &    -50.1758 &  19.969 &  0.857 &   3.961 $\pm$   0.293 & -11.429 $\pm$   0.366 &   6.01 $\pm$   0.40 &   0.23 \\ 
  9 &     13.3654 &    -50.5721 &  19.639 &  0.801 &   3.917 $\pm$   0.227 & -10.992 $\pm$   0.277 &   5.81 $\pm$   0.42 &   0.30 \\ 
 10 &     11.8523 &    -53.1060 &  19.058 &  0.000 &   4.138 $\pm$   0.177 & -11.667 $\pm$   0.220 &   6.62 $\pm$   0.82 &   0.27 \\ 
 11 &     16.8508 &    -54.0494 &  20.061 &  0.878 &   4.037 $\pm$   0.401 & -10.870 $\pm$   0.517 &   6.20 $\pm$   0.42 &   0.30 \\ 
 12 &      8.7388 &    -56.5193 &  19.299 &  0.761 &   3.338 $\pm$   0.198 & -12.359 $\pm$   0.227 &   6.43 $\pm$   0.62 &   0.75 \\ 
 13 &     11.9163 &    -59.4346 &  19.433 &  0.799 &   2.485 $\pm$   0.241 & -12.706 $\pm$   0.268 &   5.54 $\pm$   0.43 &   0.57 \\ 
 14 &     10.3915 &    -61.9728 &  19.419 &  0.900 &   2.175 $\pm$   0.260 & -12.785 $\pm$   0.221 &   5.59 $\pm$   0.44 &   0.55 \\ 
 15 &      7.5095 &    -63.8095 &  19.275 &  0.862 &   1.492 $\pm$   0.274 & -13.079 $\pm$   0.232 &   5.48 $\pm$   0.45 &   0.47 \\ 
 16 &      9.4180 &    -64.1045 &  18.997 &  0.782 &   2.062 $\pm$   0.191 & -13.123 $\pm$   0.175 &   5.62 $\pm$   0.55 &   0.22 \\ 
 17 &      6.2280 &    -68.3379 &  19.958 &  0.831 &   1.130 $\pm$   0.391 & -13.195 $\pm$   0.374 &   5.40 $\pm$   0.34 &   0.28 \\ 
 18 &    328.9156 &    -78.7878 &  19.248 &  0.915 &  -9.076 $\pm$   0.255 &  -9.534 $\pm$   0.278 &   5.46 $\pm$   0.54 &   0.26 \\ 
 19 &    269.9732 &    -69.5896 &  19.412 &  0.867 & -11.275 $\pm$   0.166 &   1.886 $\pm$   0.246 &   5.94 $\pm$   0.55 &   0.35 \\ 
 20 &    261.2942 &    -70.4575 &  20.202 &  1.020 & -11.118 $\pm$   0.377 &   2.342 $\pm$   0.473 &   6.12 $\pm$   0.92 &   0.46 \\ 
 21 &    251.1577 &    -69.8245 &  20.092 &  1.063 & -11.097 $\pm$   0.313 &   2.790 $\pm$   0.419 &   5.59 $\pm$   0.87 &   0.39 \\ 
\enddata
\end{deluxetable}

\startlongtable
\begin{deluxetable*}{rcccccccc}
\tabletypesize{\small}
\tablecaption{Candidate Stream Stars: Inner Leading Tail}
\tablecolumns{9}

\tablehead{
\colhead{No.} & \colhead{R.A. (J2016)} & \colhead{dec (J2016)} & \colhead{$G$} & \colhead{$G_{BP} - G_{RP}$} & \colhead{$\mu_\alpha \cos{\delta}$ (mas yr$^{-1}$)} & \colhead{$\mu_\delta$ (mas yr$^{-1}$)} & \colhead{Distance (kpc)} & \colhead{Probability}
}
\startdata
  1 &     13.8482 &    -29.7557 &  18.859 &  0.752 &   3.990 $\pm$   0.199 &  -6.415 $\pm$   0.279 &   8.58 $\pm$   2.16 &   0.81 \\ 
  2 &     13.6004 &    -31.1579 &  19.506 &  0.704 &   4.045 $\pm$   0.397 &  -6.617 $\pm$   0.439 &   9.40 $\pm$   1.44 &   0.79 \\ 
  3 &     12.8705 &    -32.5775 &  19.971 &  0.787 &   4.006 $\pm$   0.561 &  -6.469 $\pm$   0.649 &   8.38 $\pm$   0.77 &   0.62 \\ 
  4 &     11.4731 &    -34.0426 &  19.965 &  0.915 &   3.951 $\pm$   0.436 &  -6.905 $\pm$   0.540 &   8.71 $\pm$   0.82 &   0.62 \\ 
  5 &     12.3481 &    -34.9159 &  19.327 &  0.739 &   3.759 $\pm$   0.169 &  -7.181 $\pm$   0.348 &   8.35 $\pm$   1.13 &   0.49 \\ 
  6 &     13.7741 &    -34.9398 &  19.612 &  0.724 &   3.664 $\pm$   0.246 &  -7.390 $\pm$   0.262 &   7.54 $\pm$   0.76 &   0.48 \\ 
  7 &     11.9448 &    -35.7921 &  19.430 &  0.730 &   3.448 $\pm$   0.219 &  -7.923 $\pm$   0.376 &   8.45 $\pm$   1.32 &   0.26 \\ 
  8 &     13.9332 &    -38.0415 &  18.671 &  0.678 &   3.667 $\pm$   0.121 &  -8.461 $\pm$   0.131 &   7.94 $\pm$   1.49 &   0.26 \\ 
  9 &     11.8116 &    -39.5032 &  19.869 &  0.908 &   3.421 $\pm$   0.262 &  -8.583 $\pm$   0.306 &   7.36 $\pm$   0.60 &   0.48 \\ 
 10 &     13.1899 &    -40.2668 &  19.151 &  0.669 &   3.762 $\pm$   0.178 &  -8.480 $\pm$   0.168 &   8.01 $\pm$   1.06 &   0.38 \\ 
 11 &     13.3119 &    -40.9599 &  19.248 &  0.762 &   3.668 $\pm$   0.196 &  -8.962 $\pm$   0.204 &   7.31 $\pm$   1.01 &   0.53 \\ 
 12 &     13.5271 &    -45.7569 &  17.825 &  0.686 &   3.551 $\pm$   0.066 & -10.039 $\pm$   0.078 &   5.90 $\pm$   1.10 &   0.04 \\ 
 13 &     10.7279 &    -47.0248 &  17.987 &  0.672 &   3.147 $\pm$   0.063 & -10.483 $\pm$   0.085 &   6.20 $\pm$   0.87 &   0.22 \\ 
 14 &     11.3402 &    -48.8570 &  18.518 &  0.655 &   3.257 $\pm$   0.105 & -11.108 $\pm$   0.132 &   7.02 $\pm$   1.50 &   0.24 \\ 
 15 &      7.1764 &    -55.6117 &  19.088 &  0.848 &   2.486 $\pm$   0.158 & -12.341 $\pm$   0.207 &   6.15 $\pm$   0.68 &   0.30 \\ 
 16 &      4.9437 &    -60.5060 &  19.711 &  0.943 &   0.966 $\pm$   0.369 & -13.155 $\pm$   0.401 &   5.57 $\pm$   0.38 &   0.33 \\ 
 17 &    357.0915 &    -62.6104 &  19.892 &  1.055 &   0.054 $\pm$   1.306 & -11.432 $\pm$   1.456 &   5.37 $\pm$   0.34 &   0.26 \\ 
 18 &      1.6782 &    -63.8208 &  18.475 &  0.733 &   0.255 $\pm$   0.126 & -12.895 $\pm$   0.130 &   5.58 $\pm$   0.70 &   0.75 \\ 
 19 &    344.4885 &    -70.5853 &  18.774 &  0.829 &  -4.093 $\pm$   0.142 & -13.240 $\pm$   0.163 &   5.27 $\pm$   0.59 &   0.33 \\ 
 20 &    333.3888 &    -70.3340 &  19.559 &  0.865 &  -6.382 $\pm$   0.282 & -12.352 $\pm$   0.285 &   5.48 $\pm$   0.40 &   0.25 \\ 
 21 &    261.2942 &    -70.4575 &  20.202 &  1.020 & -11.118 $\pm$   0.377 &   2.342 $\pm$   0.473 &   6.12 $\pm$   0.92 &   0.32 \\ 
 22 &    260.9115 &    -66.2357 &  19.970 &  0.958 & -10.649 $\pm$   0.275 &   3.467 $\pm$   0.363 &   6.23 $\pm$   1.30 &   0.53 \\ 
 23 &    257.3220 &    -63.6728 &  20.198 &  1.088 &  -9.199 $\pm$   0.305 &   5.081 $\pm$   0.374 &   5.68 $\pm$   0.85 &   0.38 \\ 
\enddata
\end{deluxetable*}

\subsection{On the Origin of NGC 288}

Using a chemo-dynamical technique to assign origins to much of the
Galaxy's globular cluster population, \citet{callingham2022}
determined a high probability that NGC 288 was brought into the halo
during the Gaia-Enceladus-Sausage accretion event
\citep{belokurov2018, helmi2018}. \citet{malhan2021} and
\citet{malhan2022} demonstrated that the tidal tails of globular
clusters should show morphological and dynamical differences depending
on whether they formed {\it in situ} or whether they developed in either
cored or cuspy dark matter halos of infalling satellite
galaxies. Their modeling yielded remarkably distinct ranges for the
physical widths and velocity dispersions of stream stars one should expect in each
case. To determine whether our extended tidal tails are consistent
with an infall picture, and perhaps further elucidate the nature of
the halo of origin, we examine these quantities for the different
sub-streams in our sample.

 Using the highest probability candidates selected above, we
  measured the physical widths of the trailing tail, each of the
  leading tails, and both leading tails combined. Following
  \citet{malhan2021}, we break the tails into two to four segments and
  use our uncertainty-weighted photometric distance estimates to
  determine mean distances for each segment. We then measure the
  dispersion in the $\phi_2$ coordinates to estimate the mean
  $1\sigma$ width for each tail. Similarly, after fitting and removing
  a $\nu_{tan}$ fit (where $\nu_{tan}$ derives from the quadrature sum
  of $\mu_{\alpha}$ and $\mu_{\delta}$) we measure the dispersions in
  $\nu_{tan}$. The uncertainties are estimated by computing thousands
  of realizations of possible distributions over $\sigma_{\nu_{Tan}}$,
  given the distribution of proper motion uncertainties among our
  selected stars, and noting the upper and lower bounds where the
  $1\sigma$ distributions would allow a match to the observed data.
  Finally we compute \citet{malhan2022}'s {\it z} scores (the
  differences between the observed and simulated velocity
  dispersions relative to the uncertainties) and their associated {\it
    p} values to determine the level of agreement between our measured
  quantities and those for the small/cuspy (SCu), large/cuspy,
  small/cored (SCo), and large/cored (LCo) parent galaxies that
  \citet{malhan2022} modeled. We also compare our observed streams
  with the model stream shown in panel (b) of Figure
  \ref{models}. Table 4 lists these quantities for each of our
  extended tails individually, as well as for a combined leading tail
  that includes candidate stars from both inner and outer leading
  tails. We note that \citet{malhan2022}'s {\it in situ} and
  large/cuspy halo simulations differ significantly from our observed
  tails, with $p < 10^{-3}$ in all cases, and are therefore not
  included in the table.

 While the physical widths of the various streams are relatively
  narrow and seem to be more consistent with those found by
  \citet{malhan2021} for streams formed {\it in situ}, our computed
  $\sigma_{\nu_{Tan}}$ values suggest that NGC 288 originally formed
  and evolved in a large/cored parent galaxy. This would be consistent
  with estimates for the mass of Gaia-Enceladus ({\it e.g.}
  \citet{feuillet2020}).  It is possible that we have been too
  conservative in selecting candidate stream stars, focusing on the
  strongest perceived stream signal in either a low surface density or
  a highly contaminated field of candidates. It would take only a few
  outliers to significantly inflate the $1\sigma$ widths of the
  streams. Expected future improvements in {\it Gaia}'s proper motion
  measurements along with radial velocity confirmations should
  eventually help us to better refine the sample and improve these
  estimates. The tangential velocity dispersions agree less well with
  the model streams in Figure \ref{models} and with a small/cuspy halo
  origin, but we cannot rule either of them out at present. 

\floattable
\begin{deluxetable*}{lcccccc}
\tabletypesize{\small}
\tablecaption{Physical Widths, Tangential Velocity Dispersions, and  {\it z} Scores (with {\it
  p} values) for NGC 288's Extended Tidal Tails}
\tablecolumns{7}

\tablehead{
 \colhead{Tail} & \colhead{$1\sigma$ Width (pc)} & \colhead{$\sigma_{\nu_{Tan}}$
   (km s$^{-1}$)} & \colhead{SCu} & \colhead{SCo} & \colhead{LCo} &
 \colhead{Model Stream}
}
\startdata
Trailing & 92 & $3.04_{-0.34}^{+0.56} $ &  -1.11
(0.27) &  3.18 (0.001) & 0.89 (0.37) & 1.65 (0.10) \\
Outer Leading &    115 & $2.70_{-0.5}^{+1.2}$ & -0.82 (0.41) &  1.89
(0.06) & 0.54 (0.59) & -1.01 (0.31) \\
Inner Leading & 89 & $1.99_{-0.43}^{+1.09}$ & -1.54 (0.12) & 0.74
(0.46) & -0.74 (0.94) & -1.76 (0.08) \\
Combined Leading & 182 & $2.47_{-0.39}^{+0.75}$ & -1.59 (0.11) &  1.78
(0.08) & 0.35 (0.73) & -1.91 (0.06) \\
\enddata
\end{deluxetable*}

\section{Conclusions} \label{conclusions}

Using Pan-STARRS, DECaLS, and Gaia DR3 photometry, proper motion, and
parallax measurements we have detected both leading and trailing tidal
tails of NGC 288. At surface densities well above the background,
these tails extend at least $40\arcdeg$ and $35\arcdeg$ from the
cluster, respectively. The leading tail appears to be composed of two
or more spatially offset and kinematically distinct streams, in
agreement with modeling results both here and in the literature. While
not as clearly distinct from field stars, highly probable NGC 288
member stars along the expected trajectories of the tails extend as
far $80\arcdeg$ and $70\arcdeg$ from the cluster, respectively.

The physical widths of the tidal tails are somewhat narrower than
  one might expect if NGC 288 had originally formed in the fairly
  massive galaxy that led to the Gaia-Enceladus-Sausage event, but
  they appear to agree with the spread we see in multiple generations
  of tidal tails formed in an aspherical, LMC-perturbed Milky Way 
  potential. On the other hand, the dispersions in the tangential
  velocites of our stream candidates are indeed more consistent with
  formation of the tails in a massive, cored satellite galaxy.

Verification of stream membership will require follow-up radial
velocity measurements. If even a few of the most outlying candidates
can be confirmed as having once belonged to NGC 288, this stream will
become another particularly sensitive probe of the shape of the inner
halo potential and an important contributor to our understanding of
the influence of the LMC and other components of the Galaxy.
Disentangling the different components of NGC 288's leading tidal tail
in phase space should ultimately help us to better understand the
process of weak tidal stripping, the history of the cluster's orbit,
and the evolution of the Galactic potential over time.

\begin{acknowledgements}

  We are grateful to an anonymous referee for several very useful
  suggestions that greatly improved the scope and content of the
  manuscript. 
  
This work has made use of data from the European Space
Agency (ESA) mission {\it Gaia}
(\url{https://www.cosmos.esa.int/gaia}), processed by the {\it Gaia}
Data Processing and Analysis Consortium (DPAC,
\url{https://www.cosmos.esa.int/web/gaia/dpac/consortium}). Funding
for the DPAC has been provided by national institutions, in particular
the institutions participating in the {\it Gaia} Multilateral
Agreement.

The Pan-STARRS1 Surveys (PS1) and the PS1 public science archive have
been made possible through contributions by the Institute for
Astronomy, the University of Hawaii, the Pan-STARRS Project Office,
the Max-Planck Society and its participating institutes, the Max
Planck Institute for Astronomy, Heidelberg and the Max Planck
Institute for Extraterrestrial Physics, Garching, The Johns Hopkins
University, Durham University, the University of Edinburgh, the
Queen's University Belfast, the Harvard-Smithsonian Center for
Astrophysics, the Las Cumbres Observatory Global Telescope Network
Incorporated, the National Central University of Taiwan, the Space
Telescope Science Institute, the National Aeronautics and Space
Administration under Grant No. NNX08AR22G issued through the Planetary
Science Division of the NASA Science Mission Directorate, the National
Science Foundation Grant No. AST–1238877, the University of Maryland,
Eotvos Lorand University (ELTE), the Los Alamos National Laboratory,
and the Gordon and Betty Moore Foundation.

The DESI Legacy Imaging Surveys consist of three individual and complementary projects: the Dark Energy Camera Legacy Survey (DECaLS), the Beijing-Arizona Sky Survey (BASS), and the Mayall z-band Legacy Survey (MzLS). DECaLS, BASS and MzLS together include data obtained, respectively, at the Blanco telescope, Cerro Tololo Inter-American Observatory, NSF’s NOIRLab; the Bok telescope, Steward Observatory, University of Arizona; and the Mayall telescope, Kitt Peak National Observatory, NOIRLab. NOIRLab is operated by the Association of Universities for Research in Astronomy (AURA) under a cooperative agreement with the National Science Foundation. Pipeline processing and analyses of the data were supported by NOIRLab and the Lawrence Berkeley National Laboratory (LBNL). Legacy Surveys also uses data products from the Near-Earth Object Wide-field Infrared Survey Explorer (NEOWISE), a project of the Jet Propulsion Laboratory/California Institute of Technology, funded by the National Aeronautics and Space Administration. Legacy Surveys was supported by: the Director, Office of Science, Office of High Energy Physics of the U.S. Department of Energy; the National Energy Research Scientific Computing Center, a DOE Office of Science User Facility; the U.S. National Science Foundation, Division of Astronomical Sciences; the National Astronomical Observatories of China, the Chinese Academy of Sciences and the Chinese National Natural Science Foundation. LBNL is managed by the Regents of the University of California under contract to the U.S. Department of Energy. The complete acknowledgments can be found at https://www.legacysurvey.org/acknowledgment/.

\end{acknowledgements}

{\it Facilities:} \facility{Gaia, Pan-STARRS, DECaLS, SDSS, LaMost}

\bibliography{cjg_my_references}{}
\bibliographystyle{aasjournal}

\end{document}